\documentclass[twocolumn,tighten]{aastex6}

\usepackage{placeins}
\usepackage{natbib}
\usepackage{amsmath}
\usepackage{color}
\usepackage{enumerate}

\begin{document}

\newcommand\be{\begin{equation}}
\newcommand\en{\end{equation}}

\shorttitle{Planet-driven Spiral Arms: Formation Mechanism} 
\shortauthors{Bae \& Zhu}

\title{PLANET-DRIVEN SPIRAL ARMS IN PROTOPLANETARY DISKS: I. FORMATION MECHANISM}

\author{Jaehan Bae\altaffilmark{1,2,3} and Zhaohuan Zhu\altaffilmark{4}}

\altaffiltext{1}{Department of Terrestrial Magnetism, Carnegie Institution for Science, 5241 Broad Branch Road, NW, Washington, DC 20015, USA} 
\altaffiltext{2}{Department of Astronomy, University of Michigan, 1085 S. University Avenue, Ann Arbor, MI 48109, USA} 
\altaffiltext{3}{Rubin Fellow}
\altaffiltext{4}{Department of Physics and Astronomy, University of Nevada, Las Vegas, 4505 South Maryland Parkway, Las Vegas, NV 89154, USA}

\email{jbae@carnegiescience.edu}

\begin{abstract}

Protoplanetary disk simulations show that a single planet can excite more than one spiral arm, possibly explaining recent observations of multiple spiral arms in some systems.
In this paper, we explain the mechanism by which a planet excites multiple spiral arms in a protoplanetary disk.
Contrary to previous speculations, the formation of both primary and additional arms can be understood as a linear process when the planet mass is sufficiently small.
A planet resonantly interacts with epicyclic oscillations in the disk, launching spiral wave modes around the Lindblad resonances.
When a set of wave modes is in phase, they can constructively interfere with each other and create a spiral arm. 
More than one spiral arm can form because such constructive interference can occur for different sets of wave modes, with the exact number and launching position of spiral arms dependent on the planet mass as well as the disk temperature profile.
Non-linear effects become increasingly important as the planet mass increases, resulting in spiral arms with stronger shocks and thus larger pitch angles.
This is found in common for both primary and additional arms.
When a planet has a sufficiently large mass ($\gtrsim$ 3 thermal masses for $(h/r)_p=0.1$), only two spiral arms form interior to its orbit.
The wave modes that would form a tertiary arm for smaller mass planets merge with the primary arm.
Improvements in our understanding of the formation of spiral arms can provide crucial insights into the origin of observed spiral arms in protoplanetary disks.

\end{abstract}


\section{INTRODUCTION}
\label{sec:introduction}

A point mass perturber creates disturbances in the differentially rotating (e.g., Keplerian) background disk, which appear in the form of wakes spiraling away from the perturber.
Such spiral structures are seen in observations and/or numerical simulations of a variety of astrophysical disks, including protoplanetary disks, circumplanetary disks, and disks around various binary systems (e.g., binary stars, binary black holes).

Interestingly, numerical simulations of protoplanetary disks show that one planet can launch multiple spiral arms \citep[e.g.,][]{kley99,juhasz15,dong15,zhu15,fung15,richert15,bae16a,bae16c,bae17,lee16,dongfung17,hord17}.
Possibly supporting the idea of a single planet launching multiple spiral arms, recent high-resolution observations revealed multi-armed spirals in protoplanetary disks \citep{muto12,garufi13,grady13,benisty15,perez16,stolker16,maire17,tang17,reggiani17}, while the origin of the observed spiral arms is yet to be confirmed.

As a spiral arm steepens into a shock, its angular momentum is transferred to the background disk gas, opening a gap \citep{rafikov02b}.
When a single planet excites multiple spiral arms, it can create multiple gaps and pressure bumps in between \citep{bae17,dong17}.
The resulting structure in the gas disk can produce corresponding, enhanced features in dust emission and scattering, such that a single planet can be responsible for more than one main gap.
This mechanism possibly explains some of the multiple concentric gaps and rings seen in young protoplanetary disks \citep[e.g.,][]{alma15,andrews16,tsukagoshi16,vanboekel17,isella16,zhang16,ginski16,walsh16}.
These structures can also affect subsequent planet formation in such ringed/gapped disks by collecting solid particles preferentially in local gas pressure maxima (i.e., rings) through aerodynamic drag.

Beyond the effects on dust and gas rings and gaps, spiral shocks can transport angular momentum and dissipate energy in various astrophysical disks including protoplanetary disks \citep{goodman01,ju16,ju17,rafikov16,zhu16,ryan17}.
Not surprisingly, this capability is not limited to the primary spiral arm -- the one directly attached to the planet -- but the secondary spiral arm is also known to be able to contribute to angular momentum transport \citep{arzamasskiy17}.
If there exist additional spiral arms (e.g., tertiary, quaternary, ...) and they form through the same mechanism as the primary arm, angular momentum transport is also expected for the additional arms.

The characteristics of observed spiral arms can be used to constrain the masses of unseen planets.
As we will show in a companion paper (\citealt{baezhu17}, hereafter Paper II) the number of spiral arms varies as a function of the planet mass.
In addition, it is known that the arm-to-arm separation increases as a function of planet mass (\citealt{zhu15,fung15,lee16}, Paper II).

Despite the importance and growing number of numerical/observational studies showing multi-armed spirals in protoplanetary disks, the mechanism by which a planet excites multiple arms has not been fully understood.
In \citet[][hereafter OL02]{ogilvie02}, the formation of primary arm was explained as the result of constructive interference among a set of wave modes having different azimuthal wavenumbers.
The gravitational potential of a point mass perturber $\Phi_p$ at a position $(r, \phi)$ and time $t$ can be decomposed into a Fourier series, a sum of individual azimuthal modes having azimuthal wavenumbers $m=0, 1, 2, ..., \infty$:
$\Phi_p (r,\phi,t)= \sum_{m=0}^{\infty} \Phi_m(r) \exp[im(\phi-\Omega_p t)]$.
Through the resonance between the rigid rotation of a perturbing potential and the epicyclic motions of disk material, the $m$th Fourier component of the potential launches $m$ wave modes that are evenly spaced in azimuth \citep[][see also the review by \citealt{shu16}]{goldreich78a,goldreich78,goldreich79}.
Throughout this paper, we use $n=0,1,...,m-1$ to represent each wave mode excited by the $m$th Fourier component.
Among $m$ wave modes launched by an arbitrary $m$th azimuthal component, OL02 considered the one that originates from the perturber position (hereafter $n=0$ components; see Figure \ref{fig:spirals} for illustration).
Using a linear wave theory, OL02 calculated the phases of $n=0$ components having different azimuthal wavenumbers and confirmed with numerical simulations that these wave modes add coherently, creating a primary spiral arm. 

In this paper, we show that additional spiral arms form in a similar way as the primary arm does: through constructive interference among appropriate sets of wave modes having different $m$. 
More specifically, in the inner disk, we show that the wave modes excited at the Lindblad resonance with an azimuthal shift of $2\pi/m$ from the $n=0$ component (i.e., $n=1$ components) create a secondary spiral arm.
Similarly, a tertiary spiral arm forms via constructive interference among the wave modes excited at the Lindblad resonance with an azimuthal shift of $4\pi/m$ from the $n=0$ component (i.e., $n=2$ components), and so on and so forth when possible.
In the outer disk, it is $n=m-1$ components in each $m$th azimuthal component that form a secondary spiral arm.
As we will show throughout this paper, this mechanism explains the characteristics of spiral arms known from previous numerical simulations very well.

This paper is organized as follows.
In Section \ref{sec:background}, we first compute the phases of individual wave modes excited by a planet using a linear density wave theory and show that certain sets of wave modes can be in phase.
We then carry out a suite of two-dimensional numerical simulations in Section \ref{sec:simulations} and verify that the sets of individual wave modes predicted by the linear theory add constructively on to each other, creating spiral arms.
In Section \ref{sec:nonlinear}, we highlight some important non-linear effects, based on numerical simulations with a range of planet mass covering three orders of magnitude.
We summarize our findings and conclude in Section \ref{sec:summary}.
In Paper II, we present a parameter study varying the disk temperature and planet mass and implications of the present work.

\section{EXPECTATION FROM LINEAR WAVE THEORY}
\label{sec:background}

A rigidly rotating point mass perturber in a differentially rotating disk excites density waves at the vicinity of Lindblad resonances \citep{goldreich79}.
The perturbation driven by the $m$th azimuthal Fourier component of the perturber potential in a two-dimensional $(r, \phi)$ plane can be written as
\be
\label{eqn:perturbation}
X(r,\phi,t) = \tilde{X} (r) \exp^{i \left[ \int{k(r')} {\rm d}r + m (\phi - \Omega_p t)\right]},
\en
where $\tilde{X}(r)$ is the amplitude of the perturbation, $k(r)$ is the radial wave vector, $\Omega_p$ is the orbital frequency of the perturber, and $t$ represents time.
The radial wavenumber $k$ can be related to the azimuthal wavenumber $m$ and the background disk properties through the WKBJ dispersion relation, which can be written in a two-dimensional disk as 
\be
\label{eqn:dr}
m^2(\Omega-\Omega_p)^2 = \kappa^2 + c_s^2 k^2
\en
in the absence of self-gravity and dissipation processes.
In the dispersion relation, $\Omega$ is the local orbital angular frequency, $\kappa^2 \equiv (1/r^3) {\rm d}(r^2\Omega)^2/{\rm d}r$ is the square of the epicyclic frequency, and $c_s$ is the local sound speed.

As a wave propagates, its phase varies over radius.
The phase of a wave with an azimuthal wavenumber $m$ at an arbitrary radius $r$, $\phi_m (r)$, can be obtained by integrating ${\rm d}\phi / {\rm d}r = - k/m$:
\be
\label{eqn:phase_m}
\phi_m (r) = \phi_m(r_0) + \int_{r_0}^{r} {{\rm d}\phi \over {\rm d}r'} {\rm d}r',
\en
where $\phi_m (r_0)$ is the phase at $r=r_0$.
For a Keplerian disk ($\Omega \propto r^{-3/2}$, $\kappa = \Omega$), one can re-write the  dispersion relation in Equation (\ref{eqn:dr}) as follows:
\begin{eqnarray}
\label{eqn:dr2}
{k \over m} =  {\Omega \over c_s} \left| \left(1- {r^{3/2} \over {r_p^{3/2}}} \right)^2 - {1 \over m^2} \right|^{1/2}.
\end{eqnarray}
Inserting Equation (\ref{eqn:dr2}) into Equation (\ref{eqn:phase_m}), with ${\rm d}\phi / {\rm d}r = - k/m$, we obtain 
\be
\label{eqn:phase_m2}
\phi_m (r) = \phi_m(r_0) - \int_{r_0}^{r} {\Omega(r') \over c_s(r')} \left| \left(1- {r'^{3/2} \over {r_p^{3/2}}} \right)^2 - {1 \over m^2} \right|^{1/2} {\rm d}r'.
\en

Density waves driven by a point mass perturber launch around the Lindblad resonance $r_m^{\pm}  = (1 \pm 1/m)^{2/3} r_p$, propagating inward in the inner disk and outward in the outer disk \citep{goldreich78,goldreich79}, where $r_p$ is the radius of the perturber's circular orbit.
Far from the resonance, the phases of $m$ wave modes are 
\be
\label{eqn:phase_crest}
\phi_{m,n} (r) = -{\rm sgn}(r - r_p) {\pi \over 4m} + 2\pi {n \over m},
\en
where $n=0,1, ... , m-1$, as can be inferred from the asymptotic behavior of the Airy function (\citealt{ward86}; OL02)\footnote{One may use a phase offset term $\pi/(3m)$ in Equation (\ref{eqn:phase_crest}), which is the offset at exact Lindblad resonance locations \citep{ward86}, instead of $\pi/(4m)$. The difference between the two offset values are small ($\pi/(12m)$), especially when $m \gg 1$ modes are considered, and we find that the formation mechanism of spiral arms is not affected by the choice of the offset value (i.e., $\pi/(4m)$ vs. $\pi/(3m)$).}.
Using Equation (\ref{eqn:phase_m2}) and (\ref{eqn:phase_crest}), we now obtain the phases of individual wave modes with any given combination of $m$ and $n$:
\begin{eqnarray}
\label{eqn:phase_m3}
\phi_{m,n} (r) = &-&{\rm sgn}(r - r_p) { \pi \over 4m} + 2\pi  {n \over m}  
\nonumber\\
&-& \int_{r_m^\pm}^{r} {\Omega(r') \over c_s(r')} \left| \left(1- {r'^{3/2} \over {r_p^{3/2}}} \right)^2 - {1 \over m^2} \right|^{1/2} {\rm d}r'.
\end{eqnarray}
Throughout this paper, we call Equation (\ref{eqn:phase_m3}) the phase equation.

We note that the phase equation consists of a constant component (the first two terms) that determines the launching position of wave modes in azimuth at the Lindblad resonance, and a radially varying component (the third term) that determines how tightly the wave modes are wrapped as they propagate away from the Lindblad resonance.
The tightness of the wave modes depends on the azimuthal wavenumber $m$ as well as the background disk temperature ($c_s$) and rotation profiles ($\Omega$).
The epicyclic term is often ignored in the literature, such that the dependence of the wave propagation on $m$ is neglected. 
This may be a minor effect in many cases; however, the $m$-dependency in wave propagation is what enables the formation of multiple spiral arms by a single perturber, and therefore the epicyclic term has to be included.

\begin{figure}
\centering
\epsscale{1.18}
\plotone{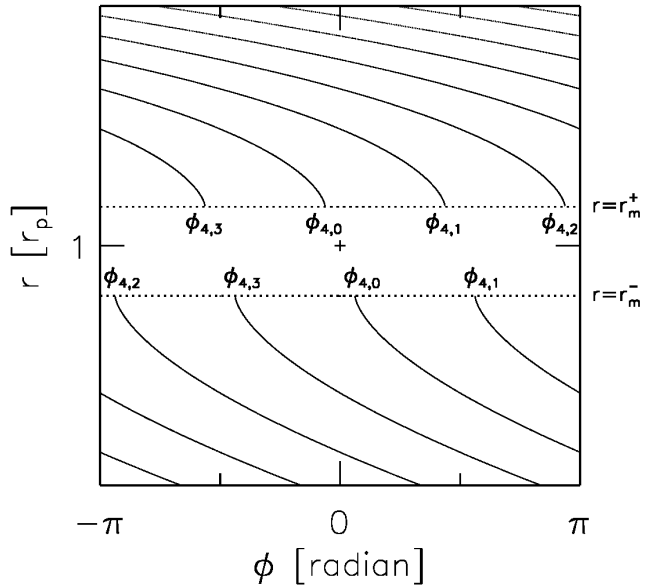}
\caption{A schematic diagram showing the phases of individual wave modes with $m=4$, as an example. The two horizontal dotted lines indicate the inner and outer Lindblad resonances $r=r_m^-$ and $r=r_m^+$, around which radii the wave modes launch. The perturber's position in the disk ($\phi=0$, $r=1r_p$) is marked with a cross symbol. The disk rotates left to right (increasing $\phi$) in this diagram.}
\label{fig:spirals}
\end{figure}

To help visualize wave excitation and propagation, we present in Figure \ref{fig:spirals} a schematic diagram showing the phases of individual wave modes with $m=4$ as an example.

\subsection{Primary Spiral Arm Formation}
\label{sec:primary_formation}

In OL02, the formation of the primary spiral arm driven by a planet in a protoplanetary disk was explained as the result of constructive interference among wave modes having different azimuthal wavenumbers.
The constructive interference considered was for $n=0$ components of each $m$th azimuthal mode.
Here, we follow OL02 and briefly summarize their findings since it will help understand the formation of additional spiral arms that will be explained in the following section.
For the example presented in this section, we adopt a temperature profile that is decreasing as a function of radius following $T \propto r^{-1/2}$ ($c_s \propto r^{-1/4}$).
In addition, we limit our attention to a Keplerian rotation profile and adopt a disk aspect ratio at $r=r_p$ of $(h/r)_p = (c_s/v_\phi)_p=0.1$ such that the sound speed is much smaller than the rotation speed.

\begin{figure*}
\centering
\epsscale{1.1}
\plotone{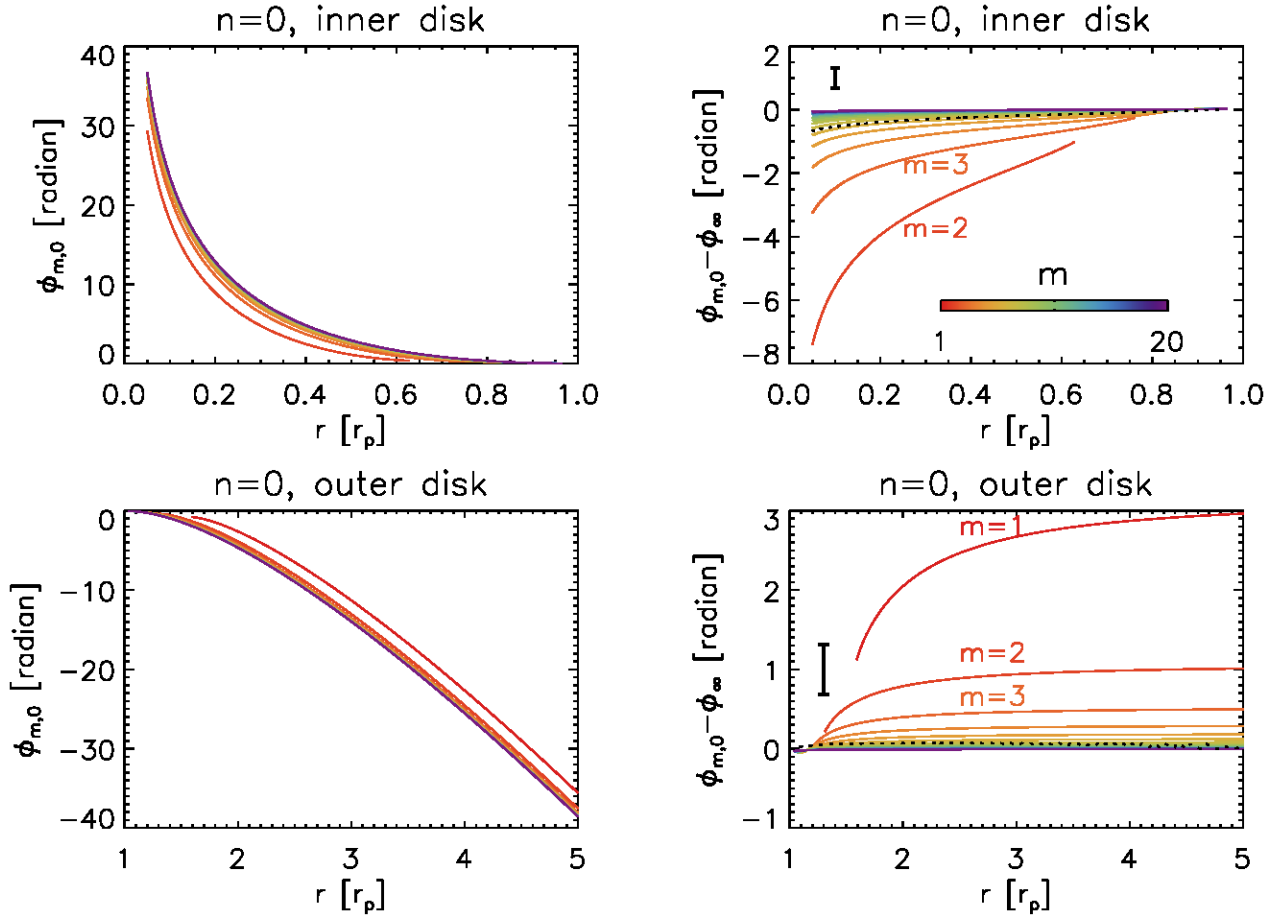}
\caption{Phases of $n=0$ wave modes with different azimuthal wavenumbers from $m=1$ (red) to 20 (purple): the left panels show the actual phase values while the right panels show the relative phase values to $m=\infty$ mode. The upper panels present phases in the inner disk ($r < r_p$) while the lower panels present phases in the outer disk ($r > r_p$). Note that there is no $m=1$ mode in the inner disk because its inner Lindblad resonance is located at $r=0$. The `I'-shaped marks in the right panels show the azimuthal width $\Delta \phi \approx 2\pi(h/r)_p$ within which different modes have to be located to participate in the constructive interference. The dotted curves in the right panels show the phase of the primary arm in the numerical simulation with full perturber potential (see Section \ref{sec:simulations}).}
\label{fig:phase_theory_prim}
\end{figure*}

For $n=0$ components, $\phi_m$ is independent of $m$ in the large $m$ limit as can be seen from the phase equation.
This implies that waves with different azimuthal wavenumbers can have the same phase so constructive interference among the waves may be possible.
In Figure \ref{fig:phase_theory_prim}, we present the phases of $n=0$ components of wave modes having azimuthal wavenumbers $m=1-20$, calculated with the phase equation.
The phases are growing positively/negatively in the inner/outer disk, meaning that these wave modes are trailing waves.
The fact that the phases become greater than $2\pi$ as they propagate to the inner disk, or smaller than $-2\pi$ in the outer disk, indicates that these wave modes can wind up multiple times before they reach the disk inner/outer boundary.

In the right panels of Figure \ref{fig:phase_theory_prim}, we present the relative phases of $n=0$ wave modes ($\phi_{m,0}$) to the phase of $m=\infty$ wave mode ($\phi_\infty$) so that the phase difference among the wave modes can be more clearly seen.
As can be seen from the figure, wave modes are nearly in phase when they launch; this is why the primary arm forms directly attached to the perturber.
However, because wave modes with a small $m$ are less tightly wound than the ones with a large $m$, as inferred from the phase equation, small $m$ modes are left behind/ahead in the inner/outer disk as they propagate.

The perturbation driven by a point mass perturber in a disk is dominated by azimuthal wavenumber $m \approx (1/2) (h/r)_p^{-1}$ \citep{goldreich80}.
In order for the wave modes with different $m$ to be coherently added, they have to be within the wave crest generated by the dominating mode, $\Delta \phi \approx 2\pi(h/r)_p$.
This azimuthal width is presented with `I'-shaped symbols in Figure \ref{fig:phase_theory_prim}.
As can be inferred from the figure, the constructive interference can fail for small $m$ wave modes far from the planet.

\begin{figure*}
\centering
\epsscale{1.1}
\plotone{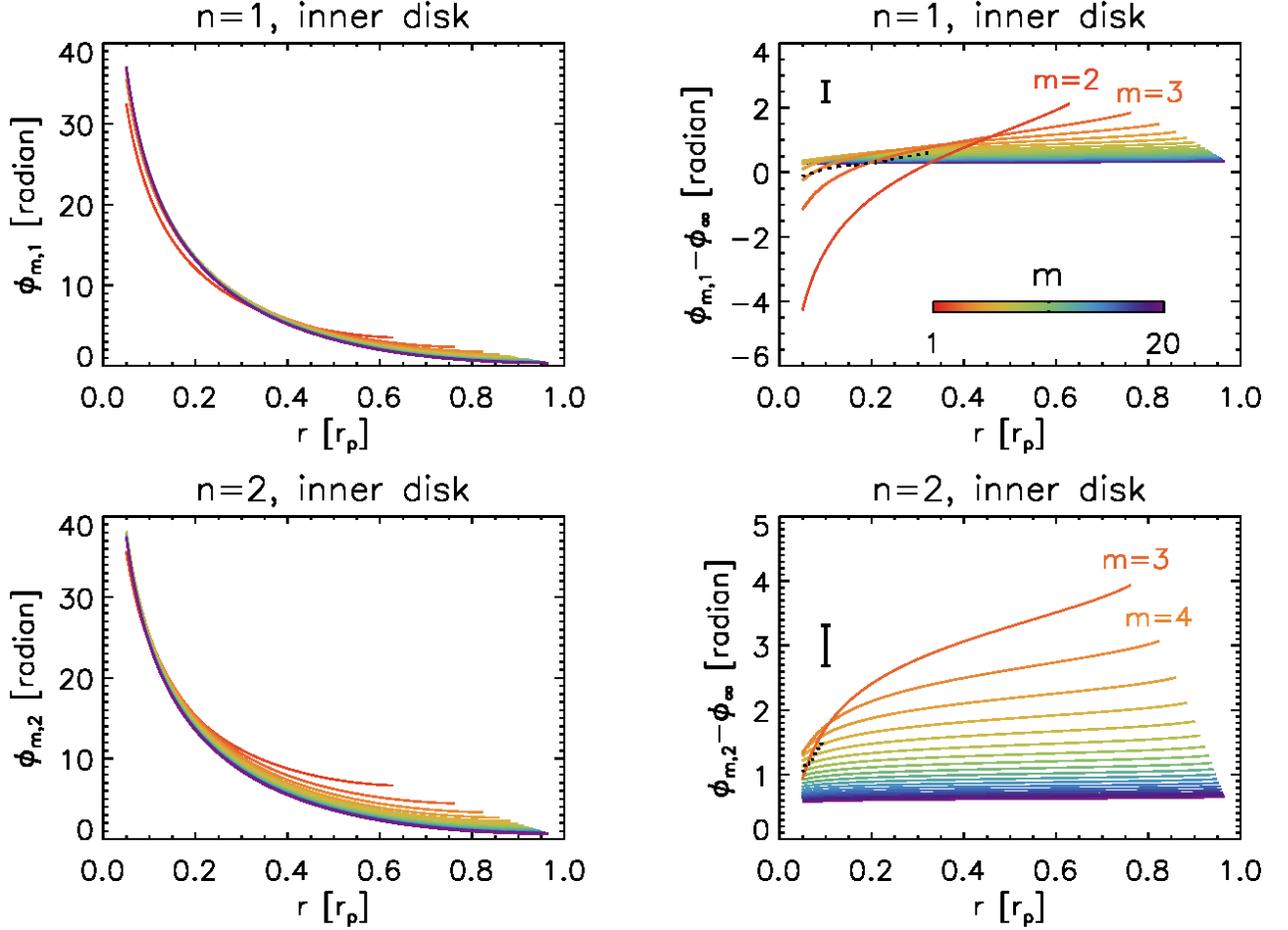}
\caption{Phases of $n=1$ and $n=2$ wave modes in the inner disk ($r < r_p$), having different azimuthal wavenumbers from $m=1$ (red) to 20 (purple): the left panels show the actual phase values while the right panels show the relative phase values to $m=\infty$ mode. Note that the shape of the phase curves for a given $m$ are identical regardless of $n$, but only the launching point is shifted in azimuth for different $n$s. The `I'-shaped marks in the right panels show the azimuthal width $\Delta \phi \approx 2\pi(h/r)_p$ within which different modes have to be located for the constructive interference to occur. The dotted curves in the right panels show the phase of the secondary and tertiary arms in the numerical simulation with full perturber potential (see Section \ref{sec:simulations}).}
\label{fig:phase_theory_inner}
\end{figure*}

\begin{figure*}
\centering
\epsscale{1.1}
\plotone{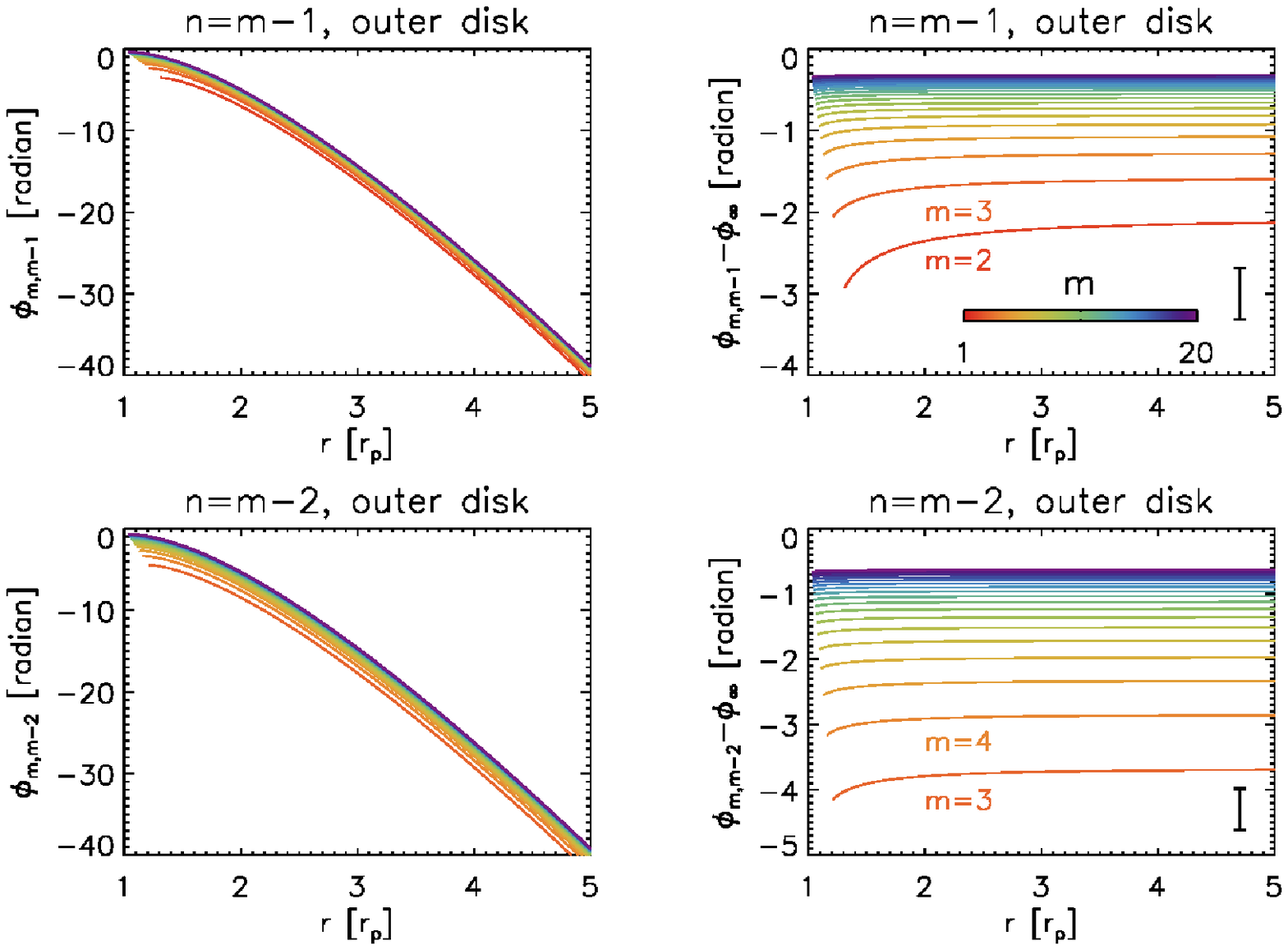}
\caption{Same as Figure \ref{fig:phase_theory_inner} but for $n=m-1$ and $n=m-2$ wave modes in the outer disk ($r > r_p$).}
\label{fig:phase_theory_outer}
\end{figure*}

\subsection{Formation of Additional Spiral Arms}

Extending the primary spiral arm formation scenario outlined in the previous section, we propose that the $n$th components of each azimuthal mode, where $n$ is now non-zero, can become in phase as they propagate and form additional spiral arms through constructive interference.
While $n=0$ components launch almost in phase as seen in Figure \ref{fig:phase_theory_prim}, other $n$ components launch with non-negligible phase differences.
For example, if one would draw $\phi_{m,1}$ for different $m$ in Figure \ref{fig:spirals}, small $m$ modes will launch with larger initial phases than large $m$ modes. 
However, small $m$ modes are less tightly wound than large $m$ modes so it is possible that small $m$ modes are caught up by large $m$ modes as the wave modes propagate.
As in Section \ref{sec:primary_formation} we compute the phases of different wave modes and examine whether or not constructive interference will be possible.
We first focus on the inner disk in Section \ref{sec:inner_disk} and then move on to the outer disk in Section \ref{sec:outer_disk}.

\subsubsection{Inner Disk}
\label{sec:inner_disk}

In Figure \ref{fig:phase_theory_inner}, we plot the phases of $n=1$ components for $m=2-20$ azimuthal modes.
As shown, small $m$ modes launch at larger azimuthal angle, but large $m$ modes catch up the small $m$ modes in phase because small $m$ modes are less tightly wound.
In this specific example, $m=2-20$ modes become in phase ($\Delta \phi \lesssim 2\pi(h/r)_p$) at $r \sim 0.3~r_p$.
The same can happen for $n=2$ components; however, small $m$ modes in this case will launch at even larger initial azimuthal angles compared with $n=1$ components, so the wave modes  have to travel further in order to become in phase.
For $n=2$ components, $m=3-20$ modes become in phase at $r \sim 0.1~r_p$.
For the disk considered here, $n > 2 $ components are unlikely to become in phase before they reach the disk inner boundary.

\subsubsection{Outer Disk}
\label{sec:outer_disk}

We now turn our attention to the outer disk.
We examine $n=m-1$ and $n=m-2$ components instead of $n=1$ and 2 components, since the wave modes considered here are trailing waves and, again, small $m$ modes are less tightly wound than large $m$ modes.
In Figure \ref{fig:phase_theory_outer}, we present the phases of $n=m-1$ and $n=m-2$ components.
As seen in the figure, the phase differences among different $m$ modes initially decrease, but remain nearly constant beyond $r \sim 3~r_p$.
It hence appears that small $m$ modes are not able to catch up to large $m$ modes.

While we present the phases out to $r=5~r_p$ only in Figure \ref{fig:phase_theory_outer}, constructive interference for $n=m-1$ and $n=m-2$ components beyond the radius is unlikely.
This can be inferred from the phase equation.
When $r \gg r_p$, the last term in the phase equation simplifies to
\begin{eqnarray}
\int_{r_m^+}^{r} {\Omega(r') \over c_s(r')} \left| \left(1- {r'^{3/2} \over {r_p^{3/2}}} \right)^2 - { 1 \over m^2} \right|^{1/2} {\rm d}r'
\nonumber\\
  \simeq  \int_{r_m^+}^{r} {\Omega(r') \over c_s(r')} \left( {r' \over {r_p}} \right) ^{3/2} {\rm d}r', 
\end{eqnarray}
and thus has no $m$ dependence.
This means that when different $m$ modes launch at different azimuthal angles and they are not in phase before $r \gg r_p$, they will not be in phase in the outer disk.
For the disk considered here, it is thus expected that only one arm forms in the outer disk through constructive interference among $n=0$ components.
On the other hand, more than one outer spiral arm can form when multiple sets of wave modes become in phase before $r \gg r_p$, which can occur in colder disks (see Paper II).

\section{NUMERICAL SIMULATIONS: VERIFYING THE LINEAR THEORY PREDICTION}
\label{sec:simulations}

In Section \ref{sec:background}, we showed that appropriate sets of wave modes having different azimuthal wavenumbers can be in phase from their launching points ($n=0$ component) or as they propagate (non-zero $n$th components).
In this section, we carry out numerical simulations to verify that constructive interference among the sets of wave modes predicted by the linear theory can indeed occur, generating  spiral arms.
We consider three models for this purpose.
\begin{enumerate}[Model 1:]
\item We carry out 20 calculations each of which includes a single $m$th Fourier-decomposed potential of a planet, where $m=1, 2, ..., 20$. We then construct a single surface density output $\Sigma$ by summing the perturbed density from each of the single mode calculation: 
$\Sigma = \Sigma_{\rm init} + \sum\limits_{m=1}^{20} {(\Sigma_m - \Sigma_{\rm init})}$, where $\Sigma_{\rm init}$ is the initial, unperturbed surface density and $\Sigma_m$ is the surface density obtained in a simulation with only the $m$th Fourier potential included.
\item We carry out one calculation in which $m=1-20$ Fourier-decomposed azimuthal components of the planet potential are included.
\item We carry out one calculation with the full planet potential.
\end{enumerate}

\begin{figure*}
\centering
\epsscale{1.18}
\plotone{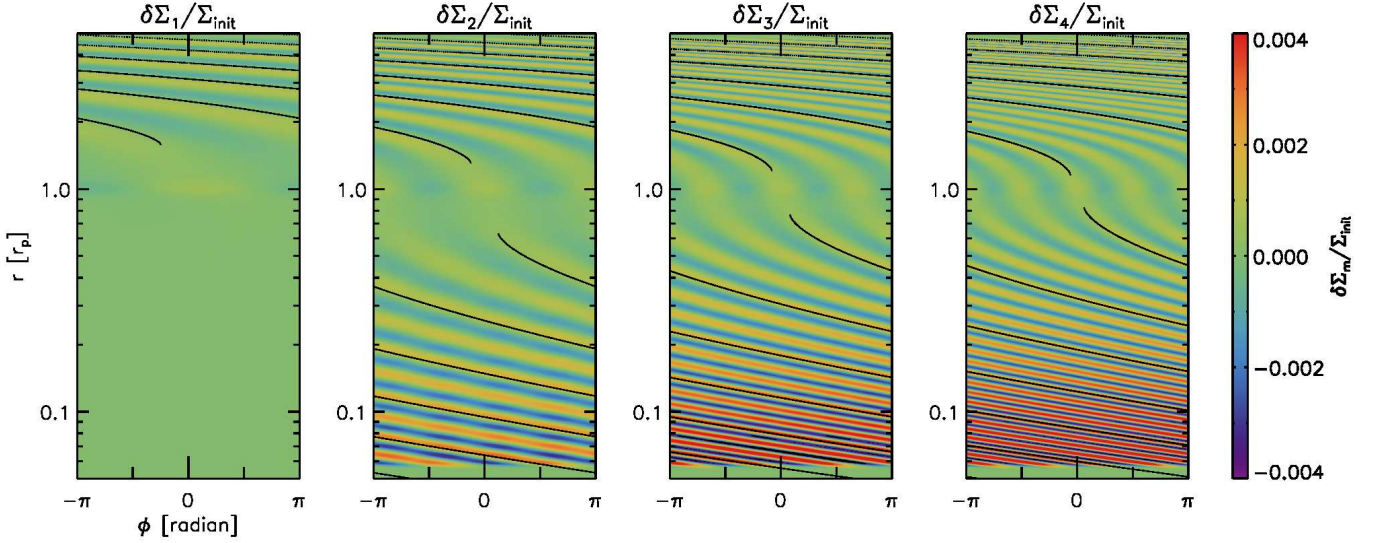}
\caption{The two-dimensional distributions of the perturbed surface density $\delta\Sigma_m/\Sigma_{\rm init}$ from individual mode runs with (left to right) $m=1$, 2, 3, and 4, where $\delta\Sigma_m = \Sigma_m - \Sigma_{\rm init}$. In each panel, the phase of $n=0$ component calculated using Equation (\ref{eqn:phase_m3}) is plotted with a black curve. Note the excellent agreement between the linear theory and numerical simulation.}
\label{fig:dens}
\end{figure*}

By comparing Model 1 and 2 with Model 3, we will be able to test whether linear addition (i.e., superposition) of individual waves explains the main features of the model with the full potential.
The comparison between Model 1 and Model 2 will allow us to examine if there exist any non-linear mode-mode interactions.
If there is no interaction between different azimuthal modes at all, we expect that Model 1 and 2 will produce identical results.
Finally, by comparing Model 2 with 3, we will be able to see the contribution from large $m$ modes ($m>20$) in generating spiral arms.

\subsection{Numerical Methods}
\label{sec:method}

We solve the hydrodynamic equations for mass and momentum conservation in the two-dimensional polar coordinates $(r,\phi)$ using FARGO 3D \citep{benitez16}:
\be\label{eqn:mass}
{\partial \Sigma \over \partial t} + \nabla \cdot (\Sigma v) = 0,
\en
\be\label{eqn:momentum}
\Sigma \left( {\partial v \over \partial t} + v \cdot \nabla v \right) = - \nabla P - \Sigma \nabla (\Phi_* + \Phi_p).
\en
In the above equations, $\Sigma$ is the surface density, $v$ is the velocity, $P=\Sigma c_s^2$ is the pressure where $c_s$ is the isothermal sound speed, $\Phi_* = -GM_*/r$ is the gravitational potential of the central star, and $\Phi_p$ is the potential of the planet.
The potential of the planet is
\be
\label{eqn:potential_full}
\Phi_p (r, \phi, t)= -{{GM_p} \over {(|{\bf{r}}-{\bf{r_p}}|^2 + s^2)^{1/2}}},
\en
where $M_p$ is the planet mass, ${\bf r}$ and ${\bf r_p}$ are the radius vectors of the center of grid cells in question and of the planet, and $s=0.6~h_p$ is the smoothing length.
In this work, we ignore the indirect term which arises due to the offset between the central star and the origin of the coordinate system.
The ``full planet potential" in Equation (\ref{eqn:potential_full}) is used for Model 3.

Assuming a circular planetary orbit, the potential in Equation (\ref{eqn:potential_full}) can be expanded into a Fourier series:
\be
\label{eqn:fourierpot}
\Phi_p (r,\phi,t)= \sum_{m=0}^{\infty} \Phi_m(r) \exp[im(\phi-\Omega_p t)].
\en
Here,
\be
\Phi_m(r) = - (2-\delta_{m0}){{GM_p} \over {2r_p}} b_{1/2}^m \left( \beta \right),
\en
where $\delta_{ij}$ is the Kronecker delta, $\beta = r/r_p$, and $b_{1/2}^m (\beta)$ is the Laplace coefficient defined as \citep{brouwer61}
\be
\label{eqn:laplace}
b_{1/2}^m (\beta) \equiv {2 \over \pi} \int_{0}^{\pi} {{\cos (m\phi)} \over {(1-2\beta \cos \phi + \beta^2 + s^2)^{1/2}}} {\rm d}\phi.
\en
The smoothing length $s$ is included in the denominator of the right-hand side of Equation (\ref{eqn:laplace}) in order for the summation of the Fourier-decomposed potential in Equation (\ref{eqn:fourierpot}) to be consistent with the full potential in Equation (\ref{eqn:potential_full}).
The Fourier-decomposed potential in Equation (\ref{eqn:fourierpot}) is used for Model 1 and 2, with $m$s chosen following the model description.

\begin{figure*}[h!]
\centering
\epsscale{1.18}
\plotone{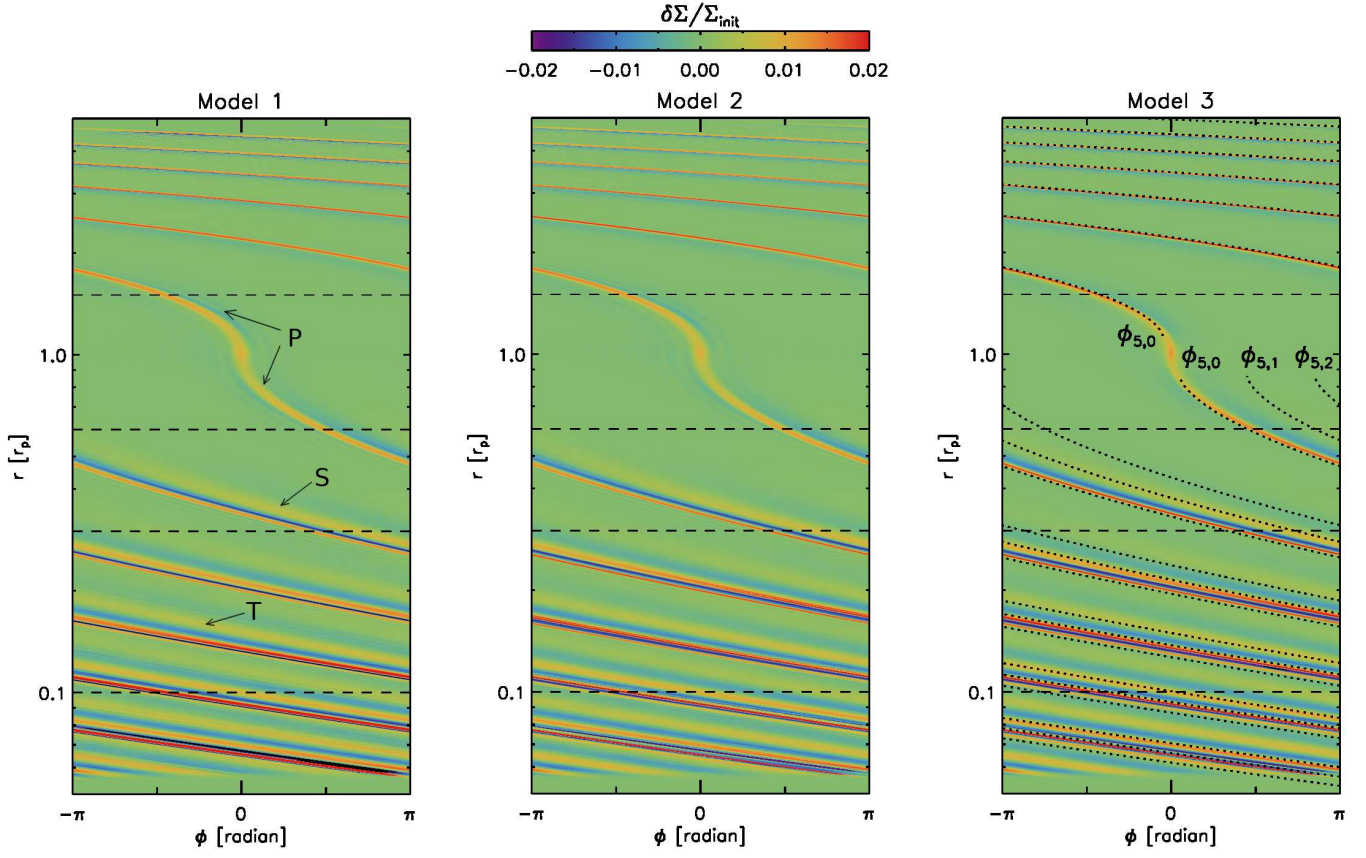}
\caption{The two-dimensional distributions of the perturbed surface density $\delta\Sigma/\Sigma_{\rm init}$ for (from left to right) Model 1, 2, and 3. The horizontal lines indicate (from top to bottom) $r=1.5, 0.6, 0.3$, and $0.1~r_p$, for which radii we present the one-dimensional perturbed density distributions along azimuth in Figure \ref{fig:dens_comp_1d}. The primary, secondary, and tertiary arms are labeled with `P', `S', and `T', respectively, in the left panel. In the right panel, the dotted curves labeled as $\phi_{5,0}$, $\phi_{5,1}$, and $\phi_{5,2}$ present the phases of the three spiral arms predicted with the linear theory.}
\label{fig:dens_comp_2d}
\end{figure*}

\begin{figure*}[h!]
\centering
\epsscale{1.15}
\plotone{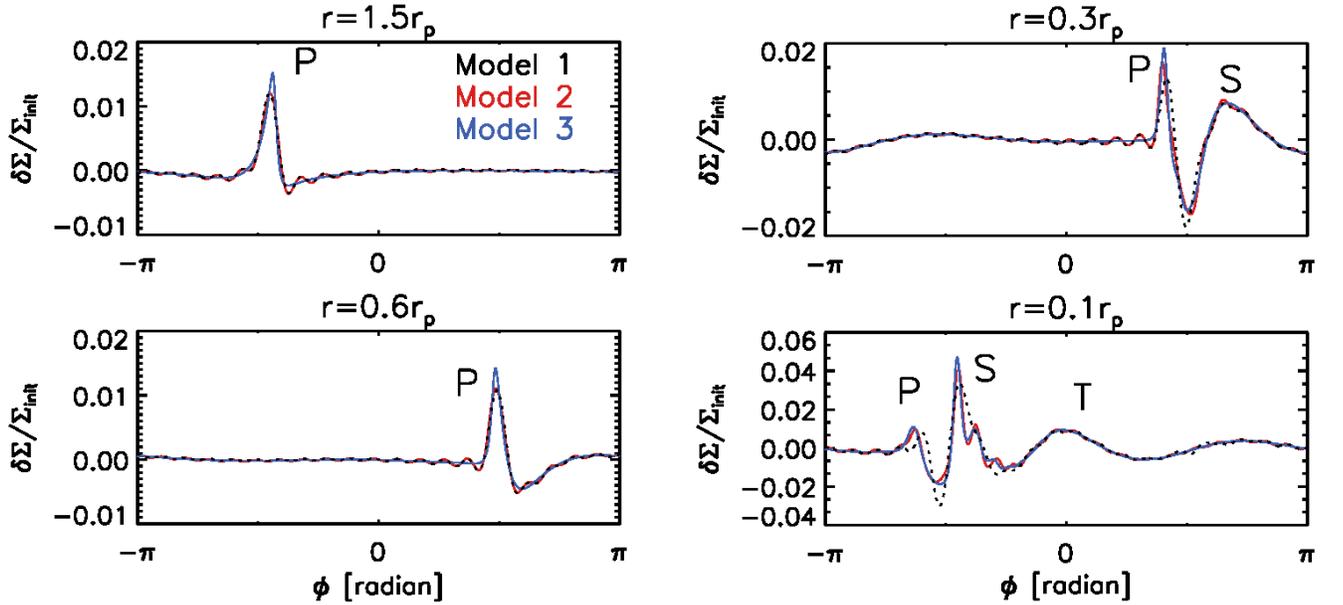}
\caption{The black dotted curves show the perturbed density distributions $\delta\Sigma/\Sigma_{\rm init}$ from Model 1, whereas the red and blue curves show results from Model 2 and 3. The primary, secondary, and tertiary arms are labeled with `P', `S', and `T', respectively.}
\label{fig:dens_comp_1d}
\end{figure*}

\begin{figure*}
\centering
\epsscale{1.1}
\plotone{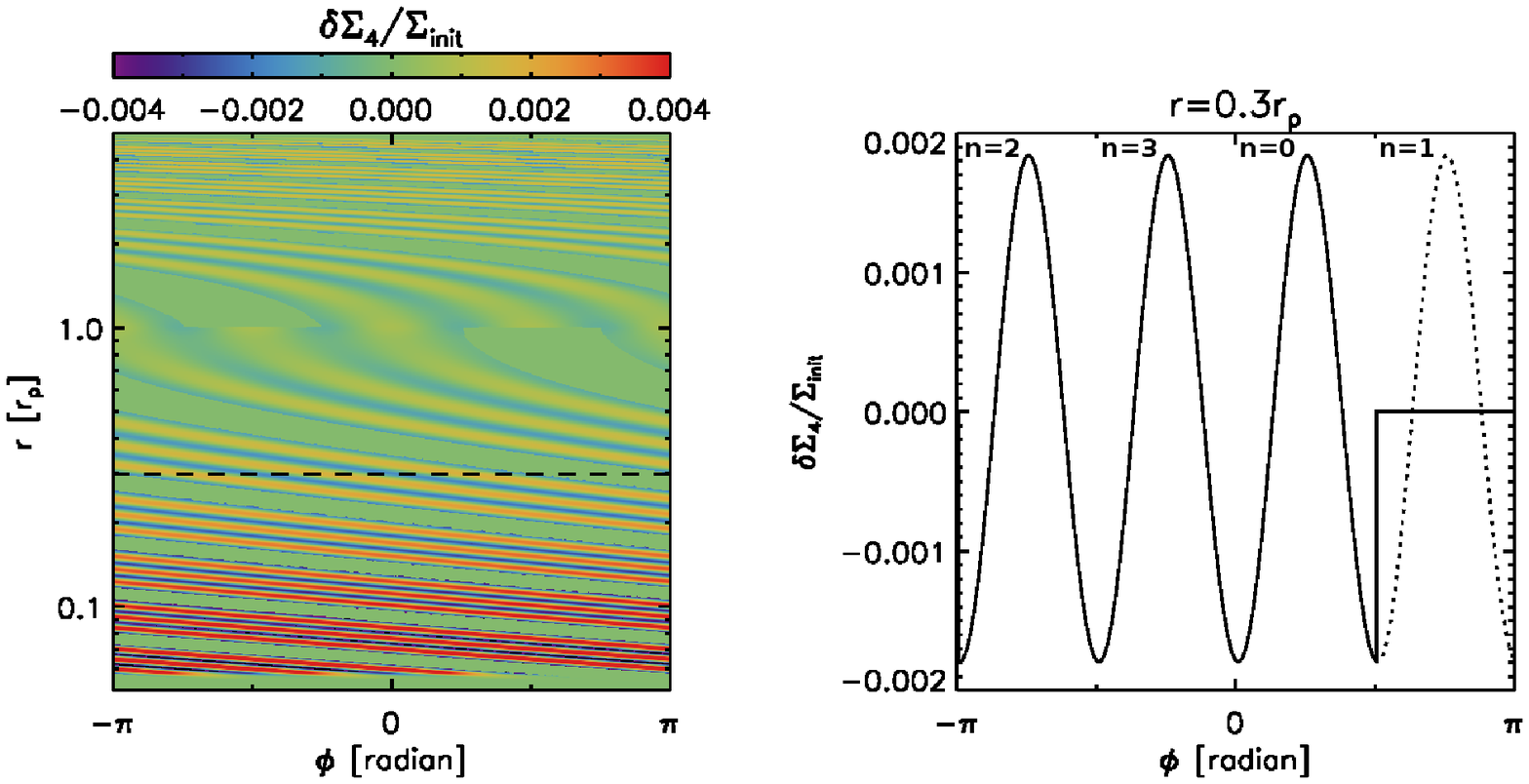}
\caption{An example showing the subtraction of selected $n$th components. In this example, $n=1$ component in the inner disk and $n=m-1$ component in the outer disk are subtracted from $m=4$ mode. (Left) The two-dimensional distribution of the perturbed surface density $\delta\Sigma_4/\Sigma_{\rm init}$ after the subtraction. The horizontal dashed line indicates $r=0.6~r_p$. (Right) $\delta\Sigma_4/\Sigma_{\rm init}$ along azimuth at $r=0.6~r_p$, before the subtraction with a dotted curve and after the subtraction with a solid curve.}
\label{fig:dens_sub}
\end{figure*}

We use the planet mass of $M_p = 0.01~M_{\rm th}$, where $M_{\rm th} \equiv c_s^3/\Omega G = M_* (h/r)_p^3$ is the so-called thermal mass \citep{lin93,goodman01}, at which the Hill radius is comparable to the disk scale height.
Assuming a solar-mass star, $0.01~M_{\rm th}$ is about 3 Earth masses. 
When $M_p \ll M_{\rm th}$, the excitation and the initial propagation of the density waves from the planet is known to be well approximated in the linear regime.
\citet{goodman01} predicts that spiral arms driven by a $0.01~M_{\rm th}$ planet steepen into shocks $\sim6$ scale heights away from the planet (see their equation 30). 
When we compare the phases of spiral arms driven by a $0.01~M_{\rm th}$ planet with the ones driven by a $0.001~M_{\rm th}$ planet in Paper II, we find that the difference in the phases of spiral arms are negligible not only within the $\pm6$ scale height regions around the planet, but in the entire disk. 
This suggests that, although spiral arms driven by a $0.01~M_{\rm th}$ planet can steepen into shocks, non-linear effects are negligible.
As shown in the following section, density waves excited by a $0.01~M_{\rm th}$ planet in numerical simulations indeed show an excellent agreement with the linear theory predictions.

Our initial disk has power-law surface density and temperature distributions:  
$\Sigma_{\rm init}(r)   =  \Sigma_p \left({r/ r_p}\right)^{-1}$ and $T(r) = T_p \left( {r /r_p} \right)^{-1/2}$,
where $\Sigma_p$ and $T_p$ are the surface density and temperature at the location of the planet $r=r_p$.
We choose $T_p$ such that $(h/r)_p = 0.1$, to be consistent with the disk model used in Section \ref{sec:background}.
The simulation domain extends from $r_{\rm in} = 0.05~r_p$ to $r_{\rm out} = 5~r_p$ in radius and from 0 to $2\pi$ in azimuth.
We adopt 4096 logarithmically-spaced grid cells in the radial direction and 5580 uniformly-spaced grid cells in the azimuthal directions, with which $\Delta r : r \Delta\phi \simeq 1:1$.
At the radial boundaries, we adopt a wave-damping zone \citep{devalborro06} to suppress wave reflection.
No kinematic viscosity is added in the simulations.

\subsection{Simulation Results}

We first present the perturbed density distributions $\delta\Sigma_m/ \Sigma_{\rm init}$, where $\delta\Sigma_m = \Sigma_m - \Sigma_{\rm init}$, from individual mode calculations (Model 1) in Figure \ref{fig:dens}.
While Figure \ref{fig:dens} includes results from $m=1 - 4$ mode runs only, we note that the discussion below applies to all individual wave mode runs with $m=1- 20$.

Most importantly, the excitation and propagation of density waves in the numerical simulations show an excellent agreement with the linear theory.
Each azimuthal component of the Fourier-decomposed potential excites $m$ wave modes at the inner and outer Lindblad resonance; $m=1$ mode does not excite any waves in the inner disk because the inner Lindblad resonance is located at $r=0$.
The perturbation from individual wave modes is $< 1\%$ over the entire simulation domain, supporting the fact that these waves are in a linear regime.
The amplitude of the perturbation in all individual mode runs increases as the waves propagate, which is also in a good agreement with the expectation for linear waves \citep{rafikov02a}.

In Figure \ref{fig:dens_comp_2d}, we display the perturbed surface density distributions $\delta\Sigma/\Sigma_{\rm init}$ for Model 1, 2, and 3.
The azimuthal distributions of $\delta\Sigma/\Sigma_{\rm init}$ at $r=1.5, 0.6, 0.3$, and $0.1~r_p$ are presented in Figure \ref{fig:dens_comp_1d} for more quantitative comparison among the models.
All three models form three spiral arms in the inner disk and one spiral arm in the outer disk.
The primary arm is directly attached to the planet, spiraling away from it.
In the inner disk, the secondary and tertiary arms excite around $r \sim 0.3~r_p$ and $r \sim 0.1~r_p$, respectively.
Note that these radial positions are in a good agreement with the predictions made based on the phase argument in Section \ref{sec:background}.
The secondary and tertiary arms become narrower in azimuth and produce stronger perturbations as they propagate, indicating that the individual waves participating in the formation of these arms become closer in phase so the constructive interference become more effective.
This is also consistent with the linear theory prediction (see e.g., Figure \ref{fig:phase_theory_inner}).

Comparing the models, we find that both Model 1 and 2 reproduce the full potential model (Model 3) fairly well.
The major difference seen in Model 3 is that the primary arm is sharper and produces a larger perturbation close to the planet (e.g., $r=0.6$ and $1.5~r_p$ in Figure \ref{fig:dens_comp_1d}).
This is because the Laplace coefficients (Equation (\ref{eqn:laplace})) that determine the strengths of the perturbation driven by individual azimuthal modes decline slowly with increasing $m$ when $\beta = r/r_p$ is close to unity.
The contribution from the azimuthal modes with $m > 20$ is therefore not negligible near $r = r_p$.
At the radius a spiral arm excites (e.g., $0.3~r_p$ for the secondary and $0.1~r_p$ for the tertiary) we see that all the three models agree with each other very well, suggesting that the excitation of additional spiral arms is a linear process.
As spiral arms propagate, however, we see Model 2 and 3 deviate from Model 1.
For example, the phase of primary arm in Model 2 and 3 is offset in phase from the primary arm phase in Model 1 at $r=0.1$ and $0.3~r_p$.
Also, at $r=0.1~r_p$ the secondary arm breaks up into finer azimuthal scales in Model 2 and 3.
This suggests that there could potentially be non-linear mode coupling \citep[e.g.,][]{lee16} even at this low level of perturbations.

To further ensure that it is the $n=0$, $n=1$, and $n=2$ components from different azimuthal modes that generate the primary, secondary, and tertiary arms, we subtract each $n$th component from Model 1 one at a time when constructing the final surface density output.
More specifically, we return the surface density associated with $n$th component in each individual azimuthal mode calculation to the unperturbed value as illustrated in Figure \ref{fig:dens_sub}.

\begin{figure*}[h!]
\centering
\epsscale{1.05}
\plotone{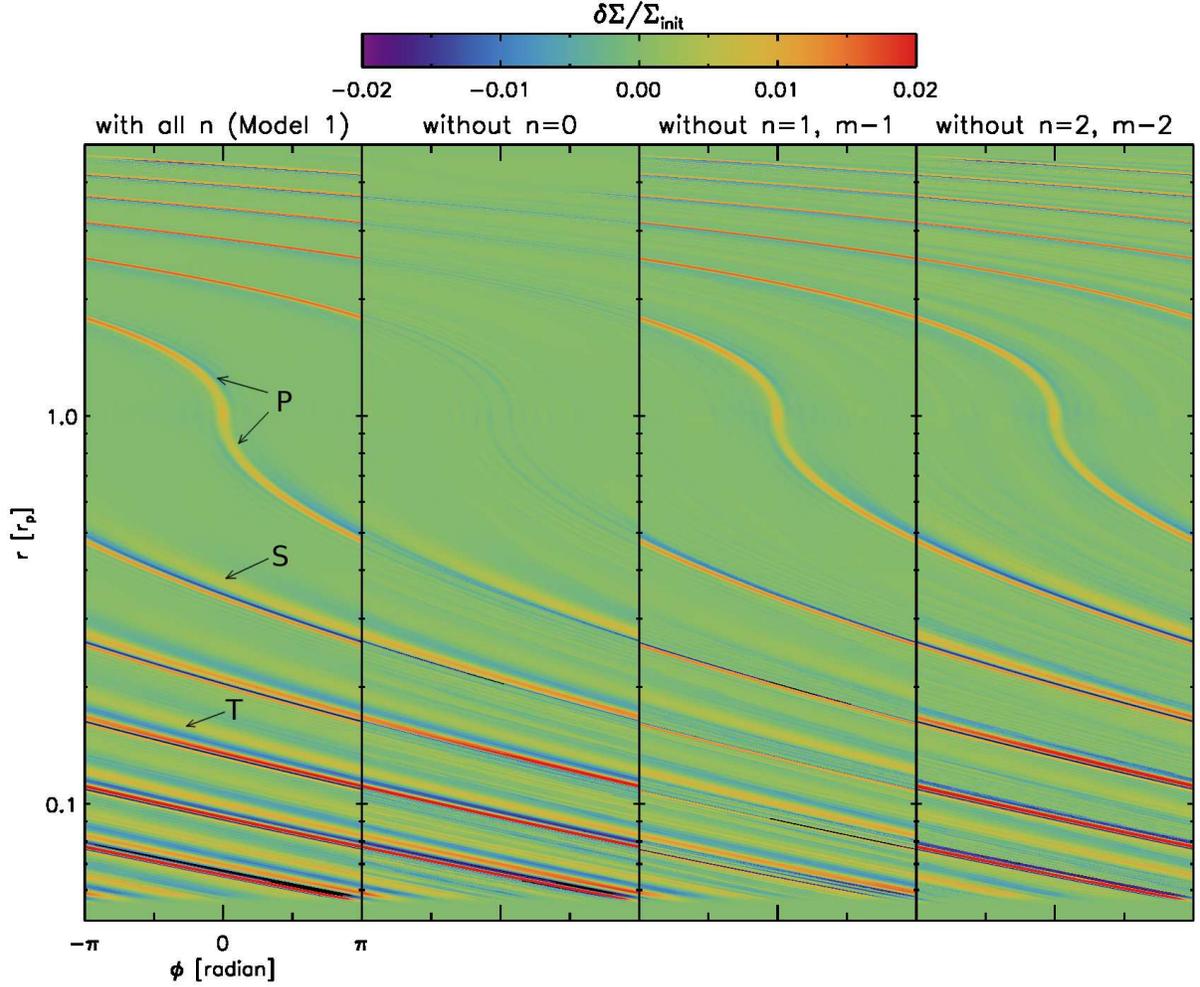}
\caption{(Left) The two-dimensional distributions of the perturbed surface density $\delta\Sigma/\Sigma_{\rm init}$ from Model 1. The primary, secondary, and tertiary arms are labeled with `P', `S', and `T', respectively. In the other three panels, the distributions of $\delta\Sigma/\Sigma_{\rm init}$ (left middle) without $n=0$ components, (right middle) without $n=1$ components at $r < r_p$ and $n=m-1$ components at $r > r_p$, and (right) without $n=2$ components at $r < r_p$ and $n=m-2$ components at $r > r_p$ are presented. Note that the primary arm does not form when the  $n=0$ components are subtracted, the secondary arm does not form when the $n=1$ components are subtracted, and the tertiary arm does not form when the $n=2$ components are subtracted.}
\label{fig:dens_sub_2d}
\end{figure*}

\begin{figure*}[h!]
\centering
\epsscale{1.1}
\plotone{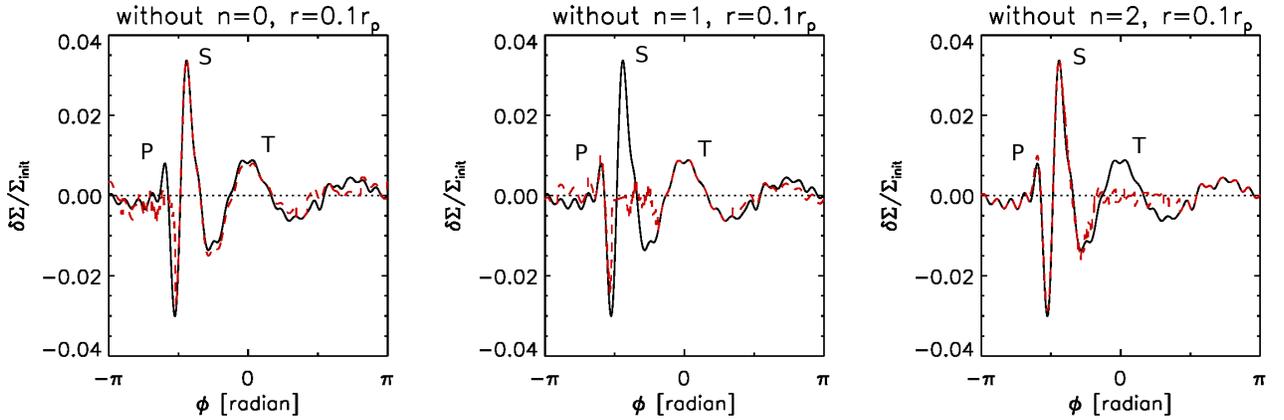}
\caption{One-dimensional plots of $\delta\Sigma/\Sigma_{\rm init}$ along azimuth at $r=0.1~r_p$. The black curves present $\delta\Sigma/\Sigma_{\rm init}$ before the subtraction, while the red curves present $\delta\Sigma/\Sigma_{\rm init}$ after the subtraction: (left) without the $n=0$ components, (middle) without the $n=1$ components, and (right) without the $n=2$ components. Note that removing certain $n$ components does not affect formation of other spiral arms. The primary, secondary, and tertiary arms are labeled with `P', `S', and `T', respectively.}
\label{fig:dens_sub_1d}
\end{figure*}

\begin{figure*}
\centering
\epsscale{1.18}
\plotone{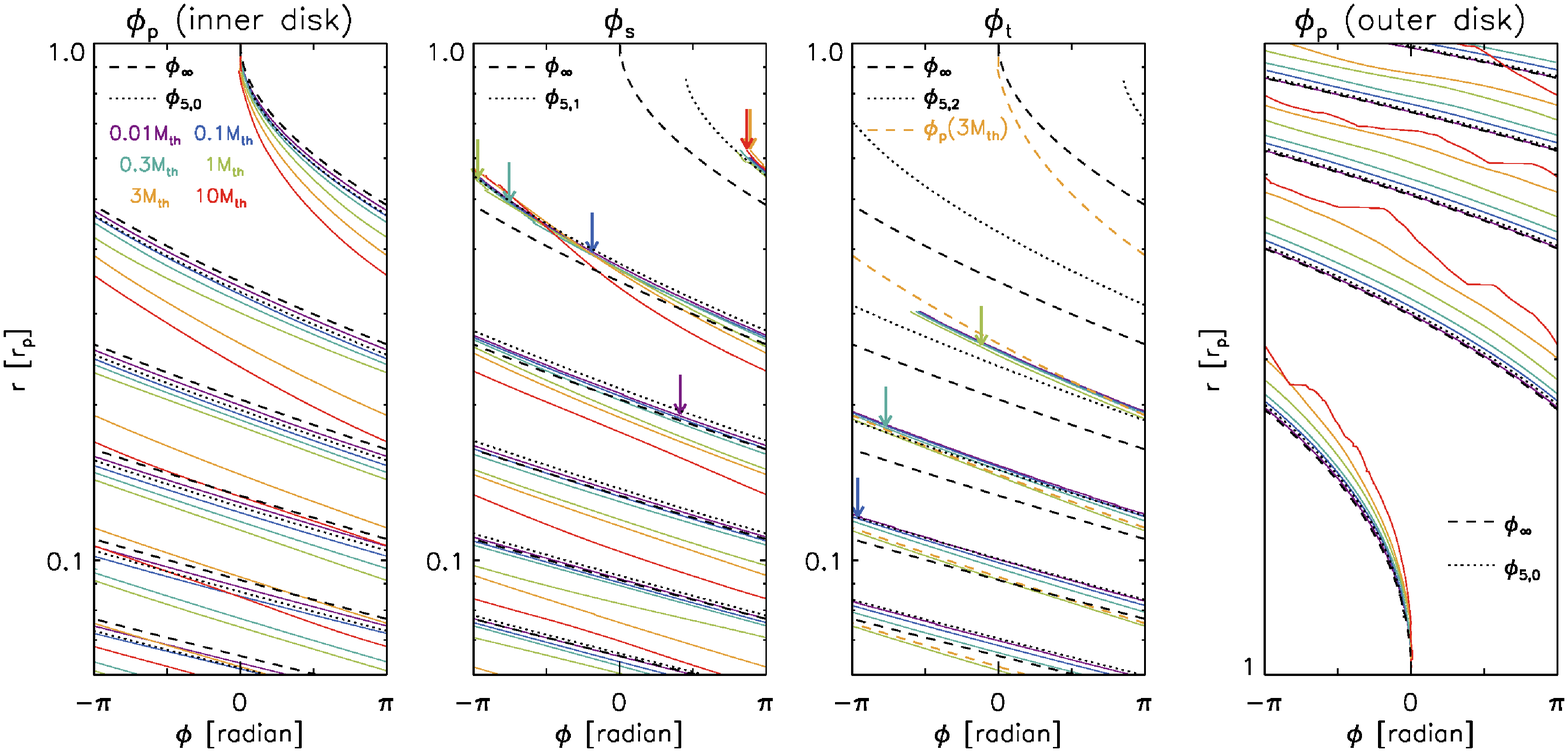}
\caption{The phases of (from left to right) the primary arm ($\phi_p$) in the inner disk, secondary arm ($\phi_s$), tertiary arm ($\phi_t$), and primary arm in the outer disk, for planet masses of 0.01, 0.1, 0.3, 1, 3, and $10~M_{\rm th}$. The dotted curves represent the phases for $m=(1/2)(h/r)_p^{-1} = 5$ mode ($\phi_{5,n}$) calculated with the phase equation, while the dashed curves represent the phase for $m=\infty$ mode ($\phi_\infty$). The arrows in the two middle panels show where each spiral arm starts to shock disk gas, diagnosed based on the potential vorticity jump. In the middle right panel, the yellow dashed curve presents the primary arm phase for the $M_p=3~M_{\rm th}$ case (see text). The distortion seen in the outer primary spiral arm for $10~M_{\rm th}$ planet is due to too strong shocks at the arm front.}
\label{fig:spiral_mass}
\end{figure*}

In Figure \ref{fig:dens_sub_2d}, we present the perturbed density distributions from Model 1 along with the ones from models without $n=0$ components, $n=1$ components ($n=m-1$ components in the outer disk), and $n=2$ components ($n=m-2$ components in the outer disk).
For more quantitative comparison among the models, we present the azimuthal distributions of the  perturbed density at $r=0.1~r_p$ in Figure \ref{fig:dens_sub_1d}.
As seen in the figures the primary arm does not form when the $n=0$ components are subtracted, the secondary arm does not form when the $n=1$ components are subtracted, and the tertiary arm does not form when the $n=2$ components are subtracted.
Also, we note that removing the $n=0$ components does not affect the secondary and tertiary arms.
Likewise, removing the $n=1$ or $n=2$ components only affects the secondary or tertiary arms.

Previous studies pointed out that a negative density perturbation appears before the secondary spiral arm forms \citep[e.g.,][]{arzamasskiy17}.
We also find such a negative density perturbation in our simulations: as shown in Figure \ref{fig:dens_comp_2d} and \ref{fig:dens_comp_1d}, the negative density perturbation just right side of the primary arm develops at $r \sim 0.6~r_p$ and deepens inward before the secondary arm launches.
In the constructive interference scenario we explain here it is obvious that, after $n=0$ components (i.e,. wave crests) form the primary arm, the wave troughs between $n=0$ and $n=1$ wave crests have to be in phase before $n=1$ wave crests become in phase to form the secondary arm. 
Similarly, the wave troughs between $n=1$ and $n=2$ wave crests become in phase before $n=2$ wave crests form the tertiary arm, and this is what forms the negative density perturbation between the secondary and tertiary arms.
In short, a negative density perturbation between spiral arms can be understood as constructive interference among wave {\it troughs}, as opposed to constructive interference among wave {\it crests} which forms a positive density perturbation (i.e., spiral arms).

Since the linear approach (i.e., superposition of individual wave modes) explains the formation and propagation of spiral arms well, we can make use of the linear wave theory to predict the phases of spiral arms.
In the right panel of Figure \ref{fig:dens_comp_2d}, we present the phases of $m=(1/2)(h/r)_p^{-1}=5$ mode for $n=0,1,$ and 2 components: $\phi_{5,0}$, $\phi_{5,1}$, and $\phi_{5,2}$.
As shown, the phases of spiral arms predicted by the linear theory agree well with the phases of spiral arms in the numerical simulation.

\section{Non-linear Evolution of Spiral Arms}
\label{sec:nonlinear}

As shown in the previous section, the linear wave theory explains the formation and propagation of spiral arms reasonably well for a sufficiently low-mass planet (i.e., $0.01~M_{\rm th}$).
As the planet mass grows, however, non-linear effects are expected to play an increasingly important role.
In order to investigate the non-linear effects, we run a set of simulations with various planet masses of $M_p = 0.1, 0.3, 1, 3$ and $10~M_{\rm th}$ (0.1, 0.3, 1, 3, and 10 Jupiter mass assuming a solar-mass star), adopting the full planet potential as in Equation (\ref{eqn:potential_full}).
All the other numerical setup except the planet mass remains the same as explained in Section \ref{sec:simulations}.

In each of the simulations we introduce the planet at the beginning of the calculation with its full mass, instead of growing the planet mass over an extended period of time.
We take this approach because planets with $M_p \gtrsim 1~M_{\rm th}$ open a gap around their orbit.
The gap edges then become unstable to the growth of the Rossby wave instability (RWI; \citealt{lovelace99,li00,li01}), launching spiral waves that have different pattern speeds from the planet-driven spiral arms.
For the planet masses considered here, we find that the RWI develops over about ten or more orbital times.
By having the full planet mass from the beginning, spiral arms launched by the planet fully develop in the entire disk well before the RWI develops.
This approach thus allows us to avoid the interference from RWI-driven spiral waves.

\begin{figure}
\centering
\epsscale{1.17}
\plotone{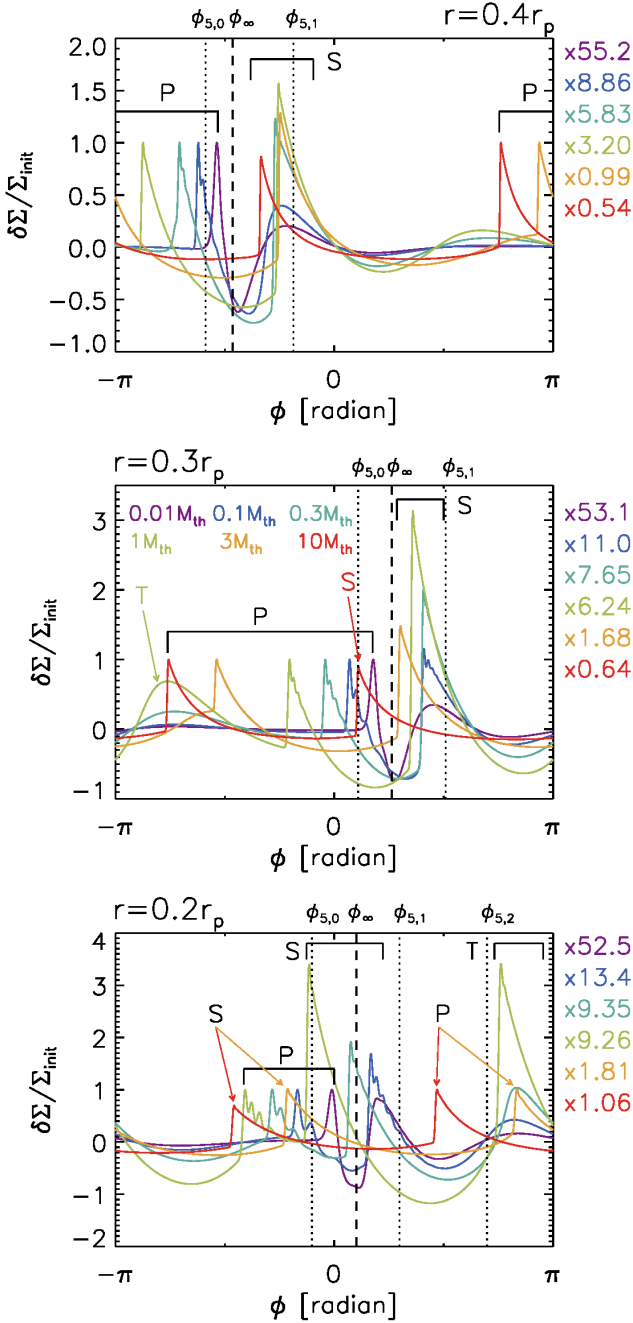}
\caption{The perturbed surface density distributions $\delta\Sigma/\Sigma_{\rm init}$ along azimuth for various planet masses at (top) $r=0.4~r_p$, (middle) $r=0.3~r_p$, and (bottom) $r=0.2~r_p$. For a visualization purpose, each curve is scaled by a factor presented on the right side of each panel such that the peak of the primary arm at the radius has $\delta\Sigma/\Sigma_{\rm init}=1$.
The primary, secondary, tertiary arms are indicated with `P', `S', and `T', respectively. The dotted vertical lines present $\phi_{5,0}$, $\phi_{5,1}$, and $\phi_{5,2}$ at each radius, while the dashed vertical line presents the phase of $m=\infty$ mode $\phi_\infty$ at the radius.}
\label{fig:phase_mass2}
\end{figure}

With the background disk profile assumed here, the planet excites two or three spiral arms in the inner disk depending on its mass.
In the outer disk, on the other hand, the planet excites only one spiral arm independently on the planet mass.
To determine the phases of spiral arms from the simulations, we find the local maximum of the density perturbation in azimuth as we follow each spiral arm along radius.
The phases of the primary, secondary, and tertiary arms for different planet masses are presented in Figure \ref{fig:spiral_mass}.
In the figure, we also indicate the radial locations at which secondary and tertiary spiral arms start to shock disk gas.
In order to diagnose the shock location, we compute the potential vorticity (PV) $\zeta \equiv ({{\nabla \times {\bf v}}) / {\Sigma}}$.
The idea is that the PV experiences a jump at the shock front \citep{li05,dong11,bae17}.
In Figure \ref{fig:phase_mass2}, we plot the perturbed surface density distributions along azimuth at some selected radii to show the level of perturbation driven by spiral arms and the morphology of spiral arm front.
We highlight some important aspects of spiral arm formation and propagation below.

{\it First}, the phases of secondary and tertiary arms during their excitation and initial propagation agree reasonably well with the linear wave theory prediction.
Looking at the phases of the secondary arms in Figure \ref{fig:spiral_mass} first, one can see that they are very closely located to each other in azimuth at $r=0.4-0.6~r_p$ for such a broad range of planet mass, and moreover follow the linear prediction ($\phi_{5,1}$) very well.
This is also clear in Figure \ref{fig:phase_mass2}: at $r=0.4~r_p$, the secondary spiral arms are located close to the linear theory prediction.
Note also that secondary arms from planets with masses of $\geq 0.3~M_{\rm th}$ have already evolved into shocks at this radius, as the steep density gradient as well as shock locations presented in Figure \ref{fig:spiral_mass} suggest.
The fact that spiral arms follow the linear theory well after they start to shock disk gas supports that shocks need to propagate some distance before they deviate from the linear theory \citep{goodman01}.
Similarly, the tertiary arms are closely located in azimuth at $r=0.15-0.2~r_p$ and the linear theory predicts the phases of the tertiary arms very well at the radii.

\begin{figure}
\centering
\epsscale{1.18}
\plotone{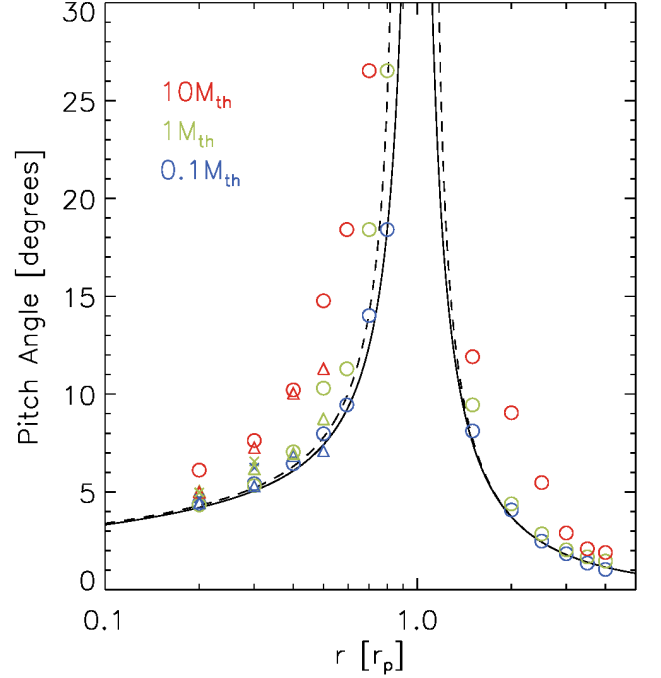}
\caption{Pitch angle of the (circle) primary, (triangle) secondary, and (cross) tertiary arm for (blue)  $0.1~M_{\rm th}$, (green) $1~M_{\rm th}$, and (red) $10~M_{\rm th}$ planets. The black solid curve presents the pitch angle calculated with $m=\infty$ in the phase equation assuming $(h/r)_p=0.1$, whereas the dashed curve presents the pitch angle calculated with $m=5$.}
\label{fig:pitch_angle}
\end{figure}

{\it Second}, spiral arms are more opened with a larger planet mass.
Spiral arms deviate from the linear theory prediction after their initial propagation, which ends sooner (i.e., at larger radii in the inner disk and at smaller radii in the outer disk) for larger planet masses (\citealt{goodman01}; see also Figure \ref{fig:spiral_mass}).
When a spiral arm non-linearly steepens into a shock it travels at a faster speed, resulting in a less tightly-wound shape than the linear theory prediction. 
The speed of the shock expansion is proportional to the amplitude of the shock, so spiral arms excited by a more massive planet propagate at faster speeds and thus appear to be more opened \citep{zhu15}.
In Figure \ref{fig:pitch_angle}, we present the measured pitch angle of spiral arms driven by 0.1, 1, $10~M_{\rm th}$ planets from numerical simulations, along with linear theory predictions.
As expected, the pitch angles measured in simulations with 1 and $10~M_{\rm th}$ planets are larger than the linear theory prediction, while the linear theory and simulation agree well with each other for a $0.1~M_{\rm th}$ planet.
Compared at a given radius, the more massive the planet is, the more opened a spiral arm is in general.
This trend is commonly seen, not only for the primary arm but also for additional arms.
We confirm this trend in disks with other $(h/r)_p$ values from the parameter study carried out in Paper II (see their Figure 5).

{\it Third}, only two spiral arms form in the inner disk for sufficiently large planet masses ($\gtrsim 3~M_{\rm th}$).
We conjecture that this is because the primary arm non-linearly propagates and merges with the wave modes that would form a tertiary arm for smaller mass planets.
In the third panel of Figure \ref{fig:spiral_mass}, we over-plot the phase of the primary arm driven by a $3~M_{\rm th}$ planet.
As shown, the primary arm becomes more opened as it non-linearly propagates to smaller radii and eventually overlaps with the linear phase of the tertiary arm at $r \sim 0.2~r_p$.
At $r=0.3~r_p$ in Figure \ref{fig:phase_mass2}, the primary arm and the tertiary arm are well separated in azimuth for a $1~M_{\rm th}$ planet.
For a $3~M_{\rm th}$ planet, on the other hand, the primary arm has a broad wing-like density enhancement on the left side of the `N'-shaped shock, which is likely formed by the waves that would constructively interfere with each other to form a tertiary arm at the azimuth.
The fact that the primary arm produces a comparable magnitude of perturbation to the secondary arm with $M_p = 3$ and $10~M_{\rm th}$ at $r=0.2$ and $0.3~r_p$, whereas the secondary arm is generally stronger for lower-mass planets at these radii because the primary arm weakens due to less efficient constructive interference, also supports the idea of tertiary arm-forming waves merging with the primary arm.

\begin{figure}
\centering
\epsscale{1.18}
\plotone{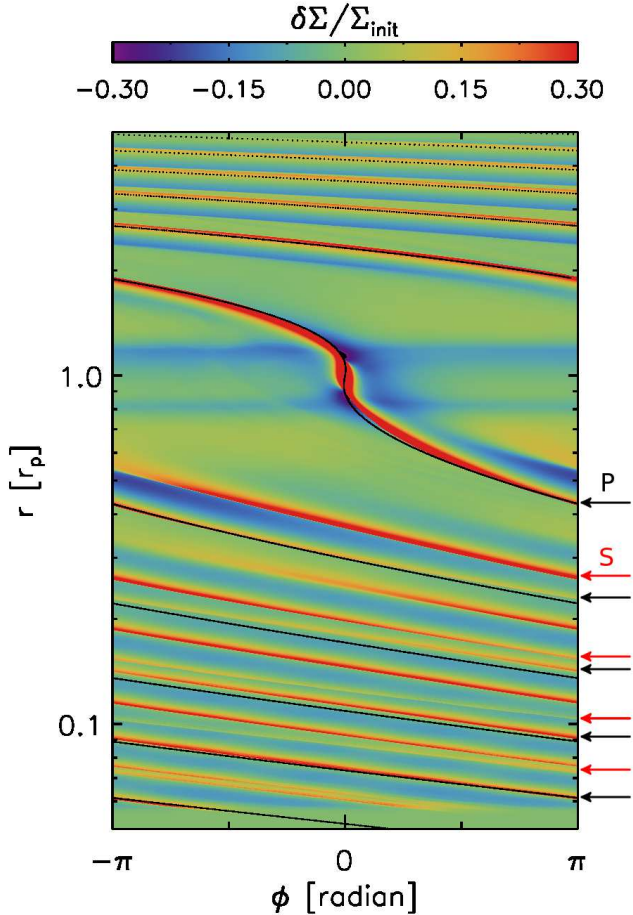}
\caption{The two-dimensional distribution of the perturbed surface density $\delta\Sigma/\Sigma_{\rm init}$ from $M_p=1M_{\rm th}$ model. The black curve presents the predicted shock front obtained with a non-linear shock expansion theory (see text). The black and red arrows on the right indicate the primary and secondary arms, respectively. Note that the primary arm deviates from the non-linear shock expansion theory prediction at $\sim0.25~r_p$ approaching to the secondary, and then returns back to the predicted phase at $\sim0.08~r_p$.}
\label{fig:shock_expansion}
\end{figure}

{\it Last}, interference between spiral arms may occur.
In Figure \ref{fig:shock_expansion} we present the two-dimensional density distribution from the $M_p = 1~M_{\rm th}$ model.
Also presented with a black curve in the figure is the predicted primary spiral arm front position based on a non-linear shock expansion theory \citep{goodman01,rafikov02a}, following the procedure detailed in Section 4 of \citet{zhu15}.
Note that the shock expansion theory predicts the primary arm phase reasonably well at $r \gtrsim 0.25~r_p$, but fails inward of the radius where the primary arm gradually approaches to the secondary arm in azimuth.
The primary arm then returns back to the predicted position at $r \lesssim 0.08~r_p$.
Interestingly, we find that this is not a transient but a long-lasting and stationary feature.
We propose that one possibility for this is the interference between the primary and secondary spiral arms.
The perturbation driven by the primary arm is expected to gradually decrease since it dissipates while propagating.
However, the primary arm gains in strength inward of $\sim 0.2~r_p$ as it approaches to the secondary arm while the secondary arm loses its strength over the same radii, possibly because some low azimuthal modes that become out of phase from the secondary arm are added to the primary arm and/or because of non-linear mode-mode interaction.
This supports the hypothesis that the deviation of the primary arm from the shock expansion theory at $0.08~r_p \lesssim r \lesssim 0.25~r_p$ is due to the interference from the secondary arm.
In Paper II, we find that interference between spiral arms in colder disks can even result in merging of the arms, presumably made possible because spiral arms in colder disks launch with smaller azimuthal separations.
While we see potential evidence of interference between spiral arms, it is unclear at the moment under which conditions such interference occurs.
Interference between spiral arms is an interesting phenomenon to study, but it is beyond the scope of the paper and thus a more thorough investigation is deferred to a future paper.

One thing that does not change regardless of planet mass (and also disk temperature; see Paper II) is that small $m$ modes are less tightly wound than large $m$ modes.
This property of waves suggests that an additional spiral arm always forms ahead of the previous arm in azimuth in the inner disk and behind the previous arm in azimuth in the outer disk.
Also, an additional spiral arm always forms farther away from the planet in radius, because the wave modes have to travel a longer distance to be in phase.

\section{SUMMARY AND CONCLUSION}
\label{sec:summary}

We have shown how a planet excites multiple spiral arms in the underlying protoplanetary disk.
Using a linear wave theory we first calculated the phases of individual wave modes excited by a planet and showed that appropriate sets of wave modes having different azimuthal wavenumbers can be in phase, from their launching points (in case of the primary arm) or as they propagate (in case of additional arms).
By carrying out a suite of two-dimensional hydrodynamic simulations, we then verified that the sets of wave modes predicted by the linear theory add constructively on to each other and form spiral arms.

As the planet mass grows, non-linear effects play an increasingly important role. 
We investigated the non-linear effects by carrying out numerical simulations with various planet masses from $1~\%$ of a thermal mass to 10 thermal masses.

Our main findings are:
\begin{enumerate}
\item The formation of spiral arms -- both primary and additional arms -- is a linear process: constructive interference among appropriate sets of wave modes having different azimuthal wavenumbers. 
\item A planet excites $m$ evenly spaced wave modes at the $m$th Lindblad resonance, where $m=1, 2, 3, ...,  \infty$ is the azimuthal wavenumber.
Among the wave modes the $n=0$ components, which are the ones originating from the planet location, add constructively on to each other and form the primary arm (Section \ref{sec:background} and \ref{sec:simulations}), confirming the mechanism presented in OL02.
\item Additional spiral arms form in a similar manner to the primary arm, but through constructive interference among non-zero $n$th components. Non-zero $n$th components excite out of phase at the Lindblad resonance, in contrast to $n=0$ components, but constructive interference among the wave modes is possible because wave propagation is dependent upon their azimuthal wavenumber (Equation \ref{eqn:phase_m3}).
\item Phases of spiral arms follow the dominating azimuthal mode with $m \approx (1/2) (h/r)_p^{-1}$ reasonably well, until they steepen into shocks and depart from the linear regime. We provide a generalized analytic formula in Equation (\ref{eqn:phase_m3}), which can be used for the primary arm but also additional arms.
\item In the outer disk, the propagation of wave modes becomes independent on azimuthal wavenumber $m$ when $r \gg r_p$. Additional spiral arms can thus form in the outer disk only if wave modes become in phase before $r \sim$ a few $\times~r_p$ (Section \ref{sec:outer_disk}).
\item Spiral arms excited by a more massive planet propagate at faster speeds and thus appear to be more opened, consistent with previous studies \citep[e.g.,][]{zhu15}. This trend is seen in common, not only for the primary arm but also for additional arms (Figure \ref{fig:pitch_angle}).
\item Only two spiral arms form for sufficiently large planet masses ($M_p \gtrsim 3~M_{\rm th}$). The wave modes that would form a tertiary arm for smaller mass planets merge with the primary arm (Figure \ref{fig:spiral_mass} and \ref{fig:phase_mass2}).
\end{enumerate}

To conclude, the multiple spiral arm formation mechanism presented in this paper can explain many characteristics of planet-driven spiral arms known from previous studies.
An obvious extension is to examine this scenario in three dimensions, particularly when the background disk is vertically stratified.
Because of vertical gravity and/or buoyancy, wave modes in such disks will behave differently as they depart from the disk midplane.
Spiral arms thus may not have a coherent vertical structure, possibly explaining the curvature of spiral arms seen in three-dimensional numerical simulations \citep[e.g.,][]{zhu15,bae16b}.
While we focused on the case in which the primary object (i.e., star) is much more massive than the perturbing companion (i.e., planet), the spiral arm formation mechanism presented here can also be applied to the systems where the companion body has a comparable mass to the primary (e.g., disks around dwarf novae or binary black holes) or a much larger mass than the primary (e.g., circumplanetary disks).
In the cases when the companion body has a mass comparable to or greater than the primary mass, it is very likely that the companion launches largely open two-armed spirals in the disk around the primary.
We present applications of the present work in Paper II and discuss whether various characteristics of observed spiral arms can be used to constrain the masses of yet unseen planets and their positions within their disks.

\acknowledgments

The authors thank the anonymous referee for a prompt report and helpful comments.
We thank Lee Hartmann, Ruobing Dong, Wing-Kit Lee, and Steve Lubow for providing valuable comments on the initial draft.
This work was supported in part by NASA grant NNX17AE31G.
ZZ acknowledges support from the National Aeronautics and Space Administration through the Astrophysics Theory Program with Grant No. NNX17AK40G and Sloan Research Fellowship. 
We acknowledge the following:  computational resources and services provided by Advanced Research Computing at the University of Michigan, Ann Arbor; the XStream computational resource, supported by the National Science Foundation Major Research Instrumentation program (ACI-1429830); the Extreme Science and Engineering Discovery Environment (XSEDE), which is supported by National Science Foundation grant number ACI-1548562; and the NASA High-End Computing (HEC) Program through the NASA Advanced Supercomputing (NAS) Division at Ames Research Center.

\end{document}